\documentstyle[aps,prl,twocolumn]{revtex}

\begin{document}
\draft

\title{An NMR-based nanostructure switch for quantum logic}

\author{John H. Reina$^{1\aleph}$,
Luis Quiroga$^{2\dagger},\;$and Neil F. Johnson$ ^{1\ddagger}$}

\address{$^{1}$Physics Department, Clarendon Laboratory, Oxford
University, Oxford, OX1 3PU, England}

\address
{$^{2}$Departamento de F\'{\i}sica, Universidad de Los Andes, Santaf\'{e}
de Bogot\'{a}, A.A. 4976, Colombia }

\twocolumn[\hsize\textwidth\columnwidth\hsize\csname
@twocolumnfalse\endcsname
\maketitle

\begin{abstract}

We propose a nanostructure switch based on nuclear magnetic
resonance (NMR) which offers reliable quantum gate operation,
an essential ingredient for building a quantum computer. The nuclear
resonance is controlled by the magic number transitions of a few-electron
quantum dot in an external magnetic field.
\end{abstract}

\pacs{PACS number(s): 73.20.Dx, 03.67.Lx, 03.67.-a, 71.10.Li, 33.25.+k}

\vskip1.6pc]
\narrowtext

Quantum superposition and entanglement are currently being
exploited to create powerful new computational algorithms in the growing
field of quantum information processing. A major question for condensed
matter physics is whether a solid-state quantum computer can ever be
built. There are at least two basic requirements which must be
met by any candidate designs: First is the ability to perform single
quantum bit (qubit) rotations as well as two-qubit controlled operations,
i.e. quantum gates. Second, the individual qubits should have a long
decoherence time.  Of utmost importance, therefore, is the identification
of a solid-state system which can be used to represent the qubits.

Nuclei with spin $\frac{1}{2}$ are natural qubits for quantum information
processing as compared to electrons\cite{Burkard}, since
they have a far longer decoherence time: indeed they have been used in
bulk liquid NMR experiments to perform some basic quantum algorithms
like those of Deutsch\cite{mosca1} and Grover\cite{mosca2}. Their
exceptionally low decoherence rates allow implementation of quantum gates
by applying a sequence of radio-frequency pulses. Nuclear spins have
already been employed in some solid-state proposals, for example that of
Kane \cite{kane} where a set of donor atoms (like P) is embedded in pure
silicon. Here, the qubit is represented by the nuclear spin of the donor
atom and single qubit and Controlled-Not (CNOT) operations might then be
achieved between neighbor nuclei by attaching electric gates on top and
between the donor atoms. Another proposal suggests controlling the
hyperfine electron-nuclear interaction via the excitation of the electron
gas in quantum Hall systems \cite{privman}. Both of these
proposals, however, require the attachment of electrodes or gates to the
sample in order to manipulate the nuclear spin qubit. Such electrodes
are likely to have an invasive effect on the coherent evolution of
the qubit, thereby destroying quantum information.

In this paper we present a new solid-state based proposal in which a
nuclear spin is coupled, not with an electron gas, but with a reduced
number of electrons in a quantum dot (QD). Our proposal avoids the
complications associated with voltage gates or electron transport by
providing an {\it all-optical} system. The nuclear resonance is controlled
by exploiting the abrupt ground-state (so-called `magic number')
transitions which arise in a few-electron QD as a function of external
magnetic field. The proposal was inspired by recent experimental results
which demonstrated the optical detection of an NMR signal in both single
QDs \cite{nmrgammon} and doped bulk semiconductors\cite{kik}.
The experimental dots were formed by interface fluctuations in GaAs/GaAlAs
quantum wells. The NMR signal from constituent Ga and As nuclei
was optically detected via excitonic recombination, exploiting
the hyperfine coupling between the electronic and nuclear systems.
Hence the underlying nuclear spins in the QD can indeed be controlled with
optical techniques, via the electron-nucleus coupling. In addition,
the experimental results of Ashoori et al. \cite{ashoori} and others, have
demonstrated that few electron (i.e. $N \geq 2$) dots can be prepared, and
their magic number transitions measured as a function of magnetic field.
The requirements for the present proposal are therefore compatible
with current experimental capabilities.

Consider a silicon-based $N-$electron QD in which a $^{13}$C impurity atom
(nuclear spin $\frac{1}{2}$) is placed at the center \cite{si}. Ordinary
silicon ($^{28}$Si) has zero nuclear spin, hence it is possible
to construct the QD such that no nuclear spins are present other than that
carried by the carbon nucleus. Since carbon is an isoelectronic impurity
in silicon, no Coulomb field is generated by this impurity. Hence
the electronic structure of the bare QD is essentially unperturbed by the
presence of the carbon atom. Suppose the quantum dot is
quasi two-dimensional (2D) and contains $N=2$ electrons.  An
external perpendicular magnetic field $B$ is applied. The
lateral confining potential in such quasi-2D QDs is typically parabolic to
a good approximation: the electrons, with effective mass $m^{\ast }$, are
confined to the $z=0$ plane with lateral confinement
$\frac{1}{2}m^{\ast}\omega _{0}^{2}\mbox{$r$}^{2}$. Repulsion
between electrons is modelled by an inverse-square interaction $\alpha
r^{-2}$ which leads to the same ground-state physics as a bare Coulomb
interaction $r^{-1}$ \cite{Quiroga1,review}: moreover, such a non-Coulomb
form may \mbox{\ actually\/} be more realistic due to the presence of
image {\mbox charges \cite{Maksym}\/}. Any many-valley effects due to the
band structure of silicon should be small and will be ignored. In
the effective mass approximation, the Hamiltonian is:
\begin{eqnarray}
H=H_{2e}+C\mathop{\displaystyle\sum}
\limits_{\nu=1}^{2}\mbox{\boldmath $I$}\cdot
\mbox{\boldmath$S$}_{\nu}\delta (\mbox{\boldmath $r$}_{\nu
})-\gamma_{n}BI_{Z}+\mathop{\displaystyle\sum}\limits_{\nu
=1}^{2}\gamma_{e}BS_{\nu,Z}\text{,}
\end{eqnarray}
where the electron-nucleus hyperfine interaction strength is given
by $C={\textstyle{8\pi \over 3}}\gamma_{e}\gamma _{n}\hbar
^{2}|\phi(z=0)|^2$, with $\phi(z=0)$ the single-electron wavefunction
evaluated at the QD plane, $\gamma _{e}$ ($\gamma _{n}$) is the electronic
(nuclear) gyromagnetic ratio and $\mbox{\boldmath $S$}_{\nu}$
($\mbox{\boldmath $I$}$) is the electron (nuclear) spin polarization. The
electron location in the QD plane is denoted by the 2D vector
$\mbox{\boldmath $r$}_{\nu}$. The first term represents the
two-electron QD with a perpendicular $B-$field, the second is the Fermi
contact hyperfine coupling of the nuclear spin with the electron spin, and
the last two terms give the nuclear and the electron-spin Zeeman energies.
Following Ref.
\cite{Quiroga1}, $H_{2e}$ split up into commuting center-of-mass (CM) motion
and relative motion ($rel$) contributions, for which exact eigenvalues and
eigenvectors can be obtained analytically. The total energy is
$E=E_{CM}+E_{rel}+E_{spin}$.  The electron-electron interaction only affects
$E_{rel}$.

The eigenstates of $H$ can be labelled as
$\left|I_{Z};N,M;n,m;S,S_{Z}\right\rangle $, where
$N\;$and $M\;$($n\;$and $m$) are the Landau and angular momentum numbers
for the CM (relative motion) coordinates; $S$ and $S_{Z}$ represent the
total electron spin and its $z$-component, while $I_{Z}$ represents the
$z$-component of the carbon nuclear spin. 
Consider the two-electron system in its ground state, i.e. $N=M=0$, $n=0$;
$m$ determines the orbital symmetry while $S=0,1$ represents the singlet
and triplet states respectively. The overall spin eigenstates have the
form $\left| I_{Z};S,S_{Z}\right\rangle _{m}$. For a given electron ground
state orbital, the spin Hamiltonian matrix elements are:

\begin{eqnarray}
\nonumber
H_{S} &={\textstyle{C
\over2}}&\mathop{\displaystyle\sum}\limits_{\nu=1}^{2}
\left\{\delta_{I_{Z},-} \delta _{I_{Z^{\prime }},+}\left\langle S^{\prime
},S_{Z}^{\prime }\right| \mbox{\boldmath $S$}_{\nu
,-}\left|S,S_{Z}\right\rangle \right. +\nonumber \\ &&
\hspace{-0.3cm}
\delta_{I_{Z},+} \delta_{I_{Z^{\prime }},-}\left\langle
S^{\prime},S_{Z}^{\prime }\right|\mbox{\boldmath $S$}_{\nu ,+}\left|
S,S_{Z}\right\rangle  +\nonumber \\ &&\left. \hspace{-0.33cm}
I_{Z}\delta _{I_{Z}I_{Z^{\prime }}}\left\langle
S^{\prime },S_{Z}^{\prime }\right| \mbox{\boldmath $S$}_{\nu ,Z}\left|
S,S_{Z}\right\rangle \right\} \left\langle m\right| \delta(\mbox{\boldmath
$r$}_{\nu})\left| m\right\rangle + \nonumber \\
&&
\hspace{-0.27cm}
H_{Zeeman}\text{,}\end{eqnarray}
where $H_{Zeeman}$ is the Zeeman term. In the presence of the $B-$field,
the low-lying energy levels all have $n=0$ and $m<0$. The relative angular
momentum $m$ of the two-electron ground state jumps in value with
increasing $B$ (see Refs. \cite{Quiroga1} and \cite{merkt}).
The particular sequence of $m$ values depends on the electron spin because
of the overall antisymmetry of the two-electron wavefunction \cite{merkt}.
For example, only odd values of $m$ arise if the $B-$field is sufficiently
large for the spin wavefunction to be symmetric (the spatial wavefunction
is then antisymmetric). These transitions, obtained analytically within
our inverse-square model, yield the same sequence of transitions as for
the Coulomb interaction. The electron-nucleus coupling depends on the
wavefunction value at the nucleus and hence on $m$. The jumps in $m$ will
therefore cause jumps in the amount of hyperfine splitting in the nuclear
spin of the carbon atom.

The nuclear spin$-$electron spin effective coupling affecting the
resonance frequency $\omega_{_{NMR}}$ of the  carbon nucleus is given by
$ \left\langle m\right| \delta (\mbox{\boldmath
$r$}_{\nu})\left|m\right\rangle\equiv\Delta _{m,\nu}$, where
\begin{equation}
\Delta_{m,\nu}=\displaystyle\int d\mbox{\boldmath $R$}
\displaystyle\int d\mbox{\boldmath $r$}\Psi _{2e}^{\ast }(\mbox{\boldmath
$R$}, \mbox{\boldmath $r$})\delta (\mbox{\boldmath $r$}_{\nu
})\Psi_{2e}(\mbox{\boldmath $R$},\mbox{\boldmath $r$}) \text{\ \ .}
\end{equation}
Here $\Psi _{2e}({\bf R},{\bf r})=\xi _{N,M}({\bf R})\zeta_{n,m}({\bf r})$
where $\xi _{N,M}({\bf R})$ ($\zeta _{n,m}({\bf r})$) is the
center-of-mass (relative) wavefunction \cite{Quiroga1}. A straightforward
calculation gives $\Delta _{m,\nu}\equiv \Delta(m)$ where
\begin{equation}
\Delta(m)={\displaystyle{1 \over \pi l^{2}2^{1+\mu_{m}}}}\text{\ \ .}
\end{equation}
Here $l=\sqrt{\hbar /m^{\ast }\omega }$ is the effective magnetic length,
the effective frequency is given by $\omega =\sqrt{\omega
_{c}^{2}+4\omega_{0}^{2}}$, $\omega _{c}=eB/m^{\ast }$ is the cyclotron
frequency. The term $\mu_{m}=\left( m^{2}+\frac{\alpha/l_{0}^{2}}{\hbar
\omega _{0}}\right) ^{\frac{1}{2}}$ absorbs the effects of the
electron-electron interaction and $l_{0}=\sqrt{\hbar /m^{\ast }\omega_{0}}$
is the oscillator length. Hence, the effective spin Hamiltonian $H_S$
(Eq. (2)) has the form
\begin{equation}
H_{S}=A(m)\left[(I_{_{+}}S_{_{-}}+I_{_{-}}S_{_{+}})+2I_{_{Z}}S_{_{Z}}\right]
-\gamma _{n}BI_{_{Z}}+\gamma_{e}BS_{_{Z}}\text{}
\end{equation}
where $A(m)=\frac{1}{2}C\Delta(m)$ represents a $B-$dependent hyperfine
coupling. We note that the first term of the hyperfine interaction in Eq. (5)
corresponds to the dynamic part responsible for nuclear-electron flip-flop
spin transitions while the second term describes the static shift of the
electronic and nuclear spin energy levels.

Electrons in the singlet state ($S$ = $0$) are uncoupled to the nucleus. In
this case, the nuclear resonance frequency is given by the undoped-QD NMR
signal
$\omega_{_{NMR,0}}=\gamma_n B$. For electron triplet states, the nuclear
resonance signal corresponds to a transition where the electron spin is
unaffected by a radio-frequency excitation pulse whereas the nuclear spin
experiences a flip. This occurs for the transition between states $\left|
-;1,-1\right\rangle $ and $\left|\Psi\right\rangle=c_1\left|
+;1,-1\right\rangle+c_2\left| -;1,0\right\rangle $. The coefficients $c_1$
and $c_2$ can be obtained analytically by diagonalizing the Hamiltonian given
in Eq. (5) \cite{JH}. Hence
\begin{eqnarray}
\nonumber
h\omega_{_{NMR}}={\displaystyle{3A(m)
\over 2}+{\displaystyle{1 \over 2}}\left( \gamma _{n}-\gamma
_{e}\right)B+}\hspace{2.0cm}
\\
{\displaystyle{1 \over 2}}\left[\left[
\displaystyle{A(m)}+\left( \gamma _{n}+\gamma _{e}\right) B\right]
^{2}+8A^{2}(m)\right] ^{\frac{1}{2}}.
\end{eqnarray}
Since $\gamma_e>>\gamma_n$, $h\omega_{_{NMR}}\approx\gamma_nB+2A(m)$ which
illustrates the dependence of the NMR signal on the effective $B-$dependent
hyperfine interaction. 

Figure 1 shows the effective coupling $\Delta(m)$ between the two-electron
gas and nucleus as a function of the ratio between the cyclotron frequency
and the harmonic oscillator frequency. (The CM is in its ground state). For
silicon, $C/l_0^2=60 MHz$. For $B$-field values where the electron ground
state is a spin singlet ($m$ even) no coupling is present. The strength of
the effective coupling decreases as the $B-$field increases due to the larger
spatial extension of the relative wavefunction at higher $m$ values, i.e. the
electron density at the centre of the dot becomes smaller. The
$B-$field provides a very sensitive control parameter for controlling the
electron-nucleus effective interaction. In particular, we note the large
abrupt variation of $\Delta(m)$ for $\frac{\omega_c}{\omega_0}\approx 2.1$
where the electron ground state is performing a transition from a
spin triplet state ($m=1$) to a spin triplet state ($m=3$). This ability to
tune the electron-nucleus coupling underlies the present proposal for an
NMR-based switch.

In the presence of infra-red (IR) radiation incident on the QD, the CM
wavefunction will be altered since the CM motion absorbs IR radiation. (The
relative motion remains unaffected in accordance with Kohn's theorem). This
allows an additional method for externally controlling the nucleus-electron
effective coupling, using optics. By considering the CM transition from the
ground state
$\left| N=0,M=0\right\rangle $ to the excited state $\left|
N=1,M=1\right\rangle$, which becomes the strongest transition in
high $B-$fields, we get the new spin-spin coupling term given by
\begin{equation}
\Delta_{_{\text{CM}}}(m)=\left({\displaystyle{1+\mu_{m} \over 2}}
\right) \Delta(m)\text{\ \ .}
\end{equation}
Hence the nuclear spin-electron spin coupling is
renormalized by the factor ${\textstyle{1+\mu _{m} \over 2}}$ in the
presence of IR{\mbox\ radiation\/}. By changing the location of the impurity
atom in the QD, the discontinuity strengths in Fig. 1 will be modified,
since the coupling is affected by  the  density of probability of the CM
wavefunction at the impurity site: future work will investigate the effect of
placing impurities away from the QD center.

Figure 2 shows the relative variation of $\omega_{_{NMR}}$ with{\mbox\
respect\/} to the undoped QD NMR signal, i.e. $\Delta \omega_{_{NMR}}=\frac
{\omega_{NMR}-
\omega_{NMR,0}}{\omega_{NMR,0}}$ (solid line) as a function of the frequency
ratio $\frac{\omega_c}{\omega_0}$. The jumps in the carbon nucleus\mbox{\
resonance\/} are abrupt, reaching 25\% in the absence of IR radiation.  This
allows a rapid tuning on and off resonance of an incident radio-frequency
pulse. The NMR signal in regions of spin-singlet states remains unaltered.
The
$B-$fields required to perform these jumps are relatively small (a few
Tesla). Moreover, the nuclear spin is being controlled by radio-frequency
pulses which are externally imposed, thereby offering a significant advantage
over schemes which need to fabricate and control electrostatic gates near to
the qubits, such as Refs. \cite{kane,privman}.\mbox{\ Illuminating\/} the QD
with IR light will shift the frequencies $\omega_{_{NMR}}$ (see dotted line
in Fig. 2) hence providing further all-optical control of the nuclear qubit.
A crucial aspect of the present proposal is the capability to manipulate
individual nuclear spins. All-optical NMR measurements in
semiconductor nanostructures\cite{nmrgammon,kik} together with local optical
probe experiments are quickly approaching such a level of finesse. 

The present proposal is not in principle limited to $N=2$ electrons:
generalizations \cite{chak,Quiroga2,johnson} of the present angular momentum
transitions arise for $N>2$. It was pointed out recently \cite{kouw} that
the  spin configurations in many-electron QDs could be explained in terms of
{\it just} two-electron singlet and triplet states. Therefore, the present
results may occur in QDs with $N>2$. In addition, by employing QDs
of different sizes, one could switch a subset of a QD array. Even if the QD
array is irregular, one may still be able to perform the
solid-state equivalent of the bulk/ensemble NMR computing recently reported
in Ref. \cite{gershenfeld}: this again represents a potential advantage of
the present scheme. Further advantages stem from the electrostatically
neutral character of the impurity atom $^{13}$C, and from the fact that the
silicon nuclei surrounding it have no nuclear spin: the carbon nuclear spin
state will be very effectively shielded from the environment and hence can be
expected to have an even longer decoherence time than the (charged)
donor nuclei in Ref. \cite{kane}, thereby offering reliable quantum gate
implementation.

Conditional quantum dynamics can be performed based on the selective driving
of spin resonances of the two impurity nuclear qubits $I_1, I_2$ (spin
$\frac{1}{2}$) in a system of two coupled QDs, separated by a distance $d$,
each containing two electrons. The QDs do not need to be identical in size. 
The orthonormal computation basis of single qubits $\left\{
\left|0\right\rangle ,\left| 1\right\rangle \right\}$ is represented by the
spin up and down of the impurity nuclei. The Hamiltonian (1) must
be modified: $H_{2e}$ must include the effects of the intra-dot and inter-dot
correlations between the two-body electron-electron interaction, and the
effects of the electron and nuclear spins of the second QD must be included.
Hence, we get an additional magic number transition as a function
of $B-$field which can be used for selective switching between dots, i.e.
since the ground state switches back and forth between pure and mixed states
\cite{simon}, the resonant frequency for transitions between the
states $\left| 0\right\rangle,$ and $\left| 1\right\rangle$ of one nuclear
spin (target qubit)  depends on the state of the other one (control qubit).
In this way, such coupled QDs can be used to generate the conditional CNOT
gate CNOT$_{ij}$$(\left|\varphi _{i}\right\rangle\left|
\varphi_{j}\right\rangle )\mapsto \left|\varphi _{i}\right\rangle
\left| \varphi _{i}\oplus \varphi_{j}\right\rangle $$\;$(where $\oplus $
denotes addition\ modulo\ 2 and the indices $i$ and $j$ refer to the control
bit and the target bit) on the qubits $\left| \varphi _{i}\right\rangle$,
and $\left|\varphi _{j}\right\rangle$. Single qubit rotations, e.g.
the Hadamard transformation $H^{T}(\left| 0\right\rangle)\mapsto
\frac{1}{\sqrt{2}}\left( \left| 0\right\rangle+\left| 1\right\rangle
\right), \;$and $H^{T}(\left|1\right\rangle
)\mapsto\frac{1}{\sqrt{2}}\left(\left| 0\right\rangle -\left|
1\right\rangle\right)$ can be performed by rotating the single nuclear
qubit via the application of an appropriate $B-$field. Note that the
presence or absence of an IR photon can also represent a qubit: hence the
present single QD system in an IR cavity can also be used to perform
two qubit gates as a result of the coupling.

In summary, we have proposed a solid-state qubit scheme which offers long
decoherence times and reliable implementation of quantum gates.
The fabrication  requirements are compatible with current experimental
capabilities. Being all-optical, rather than transport-based, the scheme
avoids the need for electrical contacts and gates.

J.H.R. and L.Q. acknowledge financial support from COLCIENCIAS. We thank Bruce Kane for drawing 
our attention to Ref. [10]. J.H.R. thanks the hospitality of the Condensed Matter Theory Group at
Universidad de Los Andes, where part of this work was carried out, and is
indebted to Helen Steers for continuous encouragement.

{\small
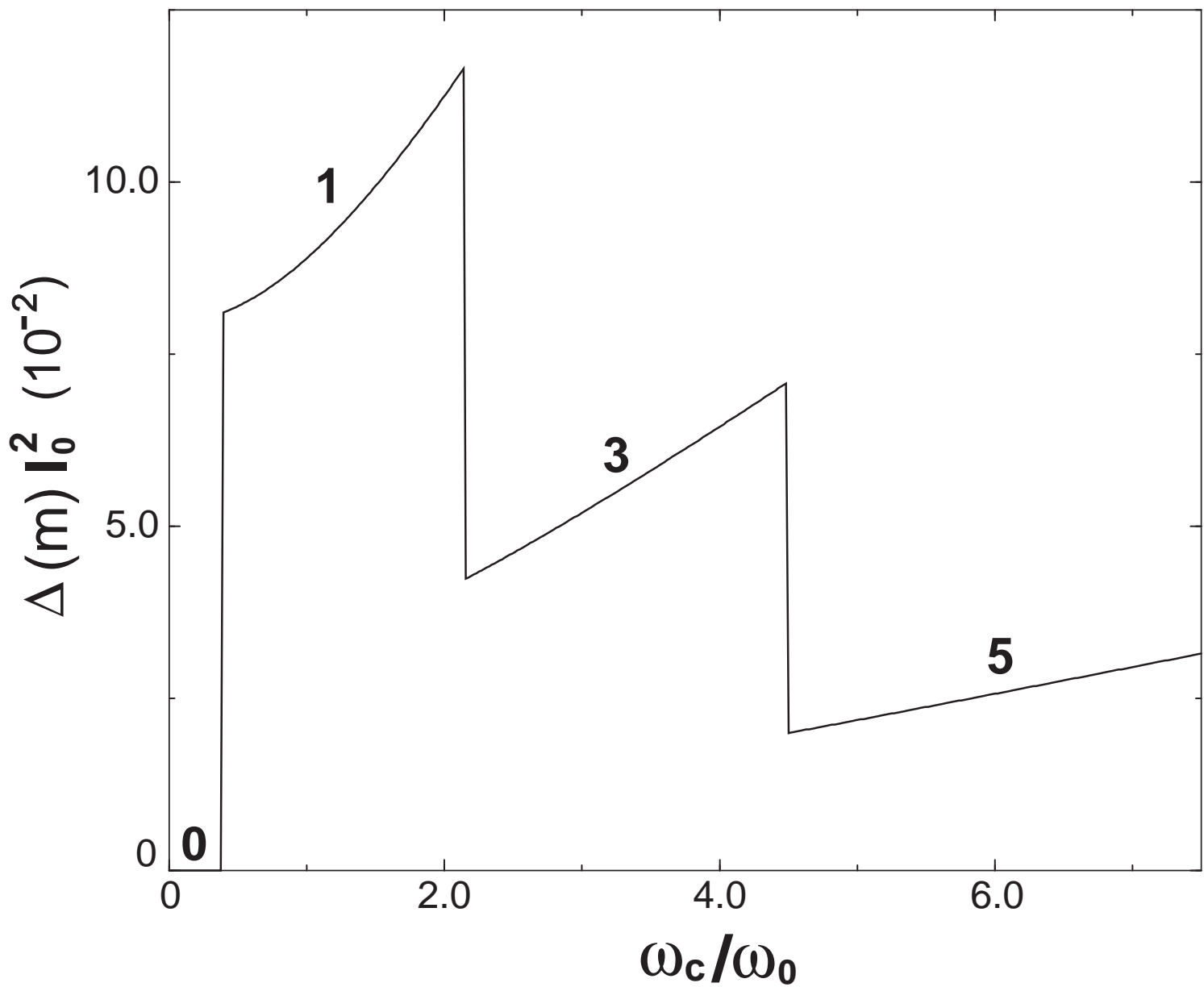
Figure 1. Variation of the electron spin -- nucleus spin effective coupling
$\Delta(m)$ as a function of $\frac{\omega_c}{\omega_0}$; $\omega_c$ is
proportional to the magnetic field and $\omega_0$ represents the QD
confining potential strength (see text). The center-of-mass motion remains
in its ground state. The electron repulsion strength is given
by $\frac{\alpha /l_{0}^{2}}{\hbar\omega
_{0}}=3.0$. The two-electron ground state undergoes transitions in
the relative angular momentum $m$. The sequence, in terms of $|m|$
and the total electron spin $S$, is given by $(|m|,S)=\{(0,0), (1,1),
(3,1),(5,1),...\}$.

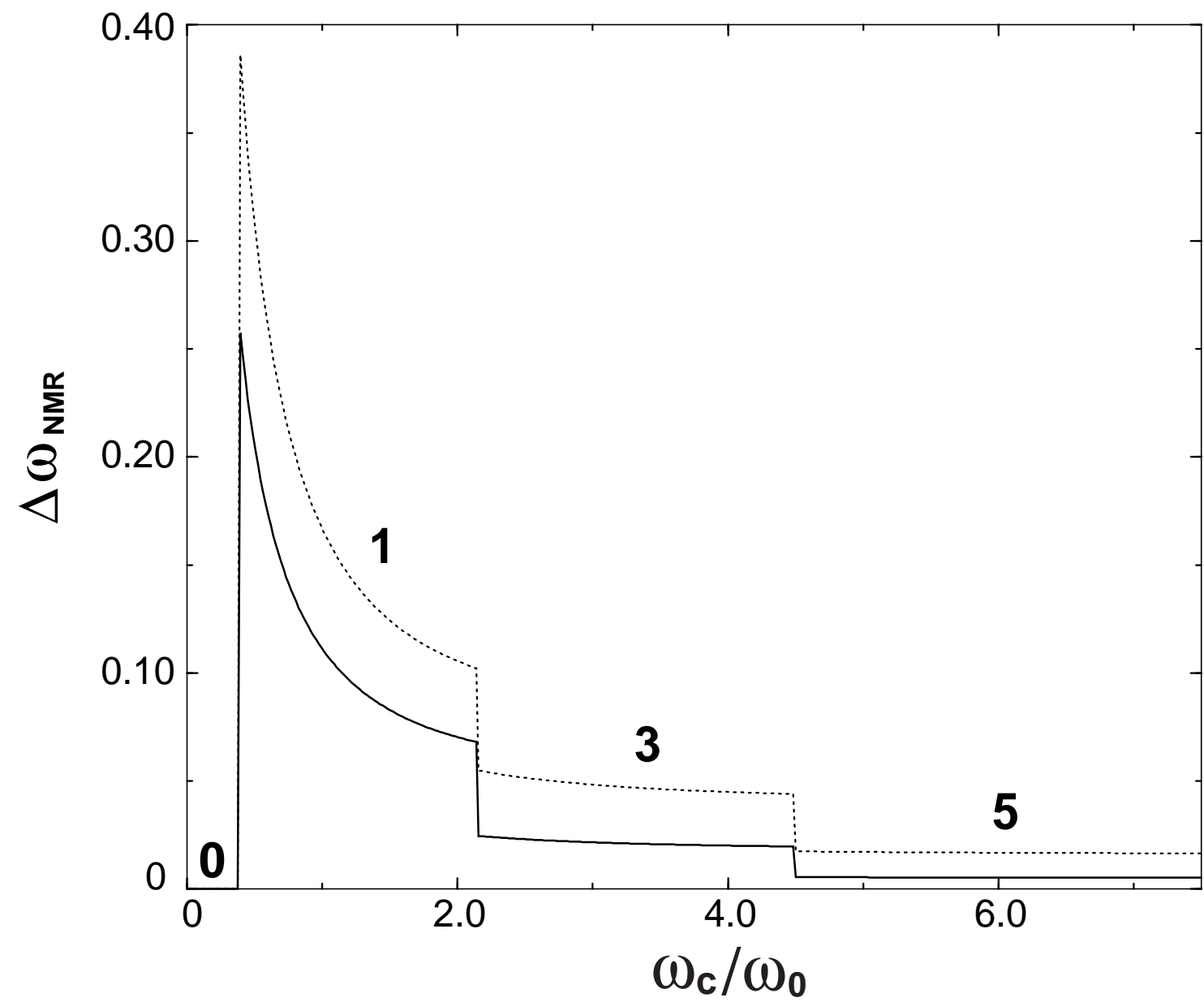
Figure 2. Relative variation of the effective nuclear magnetic resonance 
frequency  of the carbon impurity nucleus as a function
of $\frac{\omega_c}{\omega_0}$. The electron repulsion parameter is the
same as in Fig. 1. Solid line corresponds to center-of-mass in the ground
state. Dotted line corresponds to center-of-mass in the first excited state
after absorption of IR light.}


\begin{thebibliography}{99}
\vspace{-1.7cm}
\bibitem[\aleph]
:j.reina-estupinan@physics.ox.ac.uk
\bibitem[\dagger]:luis@anacaona.uniandes
.edu.co
\bibitem[\ddagger]:n.johnson@physics.ox.ac.uk

\bibitem{Burkard}
For solid state proposals using electrons as qubits, see
e.g. G. Burkard, D.
Loss, and D.P. DiVincenzo, Phys. Rev. B {\bf 59}, 2070 (1999) and references
therein.
\bibitem{mosca1}  J.A. Jones and M. Mosca, J. Chem.
Phys. {\bf 109}, 1648 (1998); I.L. Chuang, L.M.K. Vandersypen, X. Zhou, D.W.
Leung, and S. Lloyd, Nature {\bf 393}, 143 (1998).
\bibitem{mosca2}  J.A. Jones, M. Mosca, and R.H. Hansen, Nature {\bf 393},
344 (1998);  I.L. Chuang, N. Gershenfeld, and M. Kubinec, Phys.
Rev. Lett. {\bf 80}, 3408 (1998).
\bibitem{kane}  B.E. Kane, Nature
{\bf 393}, 133 (1998).
\bibitem{privman}  V. Privman, L.D. Vagner, and G.
Kventsel, Phys. Lett. A {\bf 239}, 141 (1998).
\bibitem{nmrgammon}  S.W. Brown, T.A. Kennedy, and D. Gammon, Solid
State Nuclear Mag. Res. {\bf 11}, 49 (1998).
\bibitem{kik} J.M. Kikkawa and D.D. Awschalom, Science {\bf 287},
473 (2000).
\bibitem{ashoori}  R.C. Ashoori, H.L. Stormer, J.S.
Weiner, L.N. Pfeiffer, K.W. Baldwin, and K.W. West, Phys. Rev. Lett. {\bf
71}, 613 (1993).
\bibitem{si}  There may be a natural way 
to make a quantum dot in silicon with a single C atom inside it. 
C atoms are known to act as nucleation centers for 
SiGe quantum dots: See Ref. [10]. Another possibility would be to consider an isolated $^{29}$Si
(spin 1/2
and natural abundance 4.7\%) at the center of a $^{28}$Si based QD. The
isoelectronic character of
the impurity is reinforced but possible purification procedures could be
harder to implement. 
\bibitem{schmidt}  O.G. Schmidt, S. Schieker, K. Eberl, O. Kienzle, and S. Ernst, App. Phys. Lett. {\bf 73}, 659 (1998).
\bibitem{Quiroga1}  L. Quiroga, D.R. Ardila, and N.F. Johnson, Solid
State Commun. {\bf 86}, 775 (1993).
\bibitem{review} See  N.F. Johnson, J. Phys: Condens. Matter {\bf 7}, 965
(1995) and references
therein.
\bibitem{Maksym}  L.D. Hallam, J. Weis, and P.A. Maksym, Phys. Rev. B {\bf
53},  1452 (1996).
\bibitem{merkt} M. Wagner , U. Merkt and A.V. Chaplik, Phys. Rev. B {\bf
45}, 1951 (1992).
\bibitem{JH} J.H. Reina, Clarendon Laboratory Report (unpublished).
\bibitem{chak} P.A. Maksym and T. Chakraborty, Phys. Rev. Lett. {\bf 65},
108 (1992).
\bibitem{Quiroga2}  N.F. Johnson and L. Quiroga, Phys. Rev. Lett. {\bf
74}, 4277 (1995).
\bibitem{johnson}  N.F. Johnson and L. Quiroga, J. Phys.: Condens. Matt.
{\bf 9}, 5889 (1997).
\bibitem{kouw}  S. Tarucha, D.G. Austing, Y. Tokura, W.G. van der Wiel and
L.P. Kouwenhoven, cond-mat/0001252.
\bibitem{gershenfeld}  N.A. Gershenfeld and I.L. Chuang, Science {\bf
275}, 350 (1997); D.G. Cory, A.F. Fahmy, and T.F. Havel, Proc. Natl. Acad.
Sci. USA {\bf 94}, 1634 (1997).
\bibitem{simon} S.C. Benjamin and N.F. Johnson, Phys. Rev. B {\bf 51}, 14733
(1995).

\end{thebibliography}
\end{document}